\documentclass[galaxies,article,accept,moreauthors]{Definitions/mdpi} 
\firstpage{1} 
\pubvolume{14}
\issuenum{3}
\articlenumber{53}
\pubyear{2026}
\copyrightyear{2026}
\datereceived{13 March 2026} 
\daterevised{13 April 2026} 
\dateaccepted{16 April 2026} 
\datepublished{ } 

\newcommand{\disperse}{DisPerSE}
\newcommand{\filfinder}{FilFinder}
\newcommand{\fildreams}{FilDReaMS}

\Title{Interstellar Filament Detection and Characterization: \linebreak Methods and Implications for Studies of the Magnetized Interstellar Medium}

\Author{Dana Alina \orcidA{}} 

\AuthorNames{Dana Alina}

\address[1]{%
Physics Department, School of Sciences and Humanities, Nazarbayev University. Kabanbay batyr ave, 53, \linebreak Astana 010000, Kazakhstan; dana.alina@nu.edu.kz}

\abstract{Filamentary structures are ubiquitous in the interstellar medium and play a key role in the evolution of molecular clouds and star formation. Their morphology and relative orientation with respect to magnetic fields have been widely used as a diagnostic of magnetohydrodynamic processes, turbulence, and gravitational accretion. In recent years, the growing availability of large continuum, spectroscopic, and polarization data stimulated the development of various filament detection techniques. 
In this review, we present a systematic overview of filament detection methods applied to observations of the interstellar medium. We classify the existing approaches into methodological categories, discuss underlying principles, illustrate their application on a same observational field, discuss limitations and advantages, in particular with respect to the studies of the relative alignment between magnetic fields and filaments. We conclude with presenting a point of view on the perspectives for filament studies in the era of ever-growing astronomical data volume.}


\begin{document}


\section{Introduction}

Filamentary structures are a ubiquitous component of the interstellar medium (ISM) 
 and are observed over a wide range of physical scales and environments, from diffuse atomic gas to dense molecular clouds \citep{andre2010,molinari2010,menshchikov2010}. 
The term ``filament'' is used in different fields of astrophysics to describe elongated structures, from cosmic web \citep{bond1996} 
to solar \mbox{filaments \citep{yuan2011}.} Within the ISM, filaments are observed both in atomic (HI) and molecular gas, probing different density regimes.  In this review, we mainly consider filaments in the molecular gas regimes, with an emphasis on their role in star-forming regions. We also acknowledge that a number of studies have developed filament detection methods in the context of the cosmic web, and later have found successful application in studies of the ISM \citep{sousbie2011}.
Numerical simulations of turbulent flows showed that elongated density enhancements arise naturally as a consequence of turbulent compression and shear, both in purely hydrodynamic and magnetohydrodynamic regimes \citep{padoan2001,heitsch2008,federrath2010,hennebelle2012}.
Over the past fifteen years, advances in observational sensitivity, angular resolution, and sky coverage have revealed filamentary structure as a dominant morphological element of the ISM rather than an exception associated with specific regions or evolutionary stages \citep{andre2010,planck2016-xxxv}.
The widespread observational presence and convergence with theoretical results indicate that filament formation is a fundamental outcome of ISM dynamics. The physical interpretation of processes in which filaments are involved depends critically on how they are identified and characterised.

The ISM is magnetised with magnetic fields dynamically coupled to the gas over a broad range of densities \citep{crutcher2012}. Observations across multiple wavelength, including optical and near-infrared stellar polarimetry as well as submillimeter dust polarization consistently reveal ordered magnetic field structures \citep{goodman1995,ward-thompson2000,planck2014-xix,ward-thompson2017}. These measurements establish magnetic fields as a dynamically relevant component of the ISM rather than a passive tracer. Thus, physical interpretation of filamentary structures must therefore account for the presence of magnetic fields. In addition, filamentary structures are observed over a wide range of spatial scales, from several parsecs down to sub-parsec and core scales, both in diffuse molecular clouds and in dense star-forming environments \citep{andre2010,molinari2010,arzoumanian2011,hacar2013}.

In the last decade, a significant number of studies have revealed a systematic relationship between filamentary structures and the orientation of the magnetic field \citep{planck2016-xxxiii,soler2017velac}. Observations from \textit{Planck} 
 have shown that the relative orientation between filaments and magnetic fields depends on column density and environment, with a global transition from parallel to perpendicular relative orientation as density increases \citep{planck2016-xxxv}.  
However, it remains unclear how the balance between turbulence, gravity, and magnetic fields varies with evolutionary stage and spatial scale from filamentary clouds to dense cores \citep{pattle2017,alina2019,hull2019,hennebelle2019,pillai2020,sanhueza2021}.
These observational results provide strong evidence that filaments are shaped, at least in part, by magnetohydrodynamic processes.

\textls[-15]{A wide range of filament identification techniques has been developed along with their observational discovery. These methods can be broadly classified into several methodological groups, including topological skeleton extraction (Discrete Persistent Structures Extractor, {\disperse} \cite{sousbie2011}, {\filfinder} \cite{koch2015}), multi-scale decomposition (getfilaments  \cite{menshchikov2013}), morphological segmentation ({\filfinder} \cite{koch2015}), template matching (Rolling Hough Transform, RHT \cite{clark2014}; Template Matching, TM \cite{juvela2016}; Filament Detection \& Reconstruction at Multiple Scales, {\fildreams} \cite{carriere2022fildreams}), and machine learning-based approaches \citep{alina2022malefisenta,zavagno2023,umetaliev2025}. Different groups probe different aspects of filamentary structures, such as connectivity, contrast, width, and orientation. Therefore, each group is appropriate for addressing different scientific questions. The choice of method should thus be guided by the underlying physical motivation.}

\textls[-15]{Among these approaches, methods designed to quantify the relative orientation between filamentary structures and magnetic fields are particularly well-suited for investigating magnetohydrodynamic (MHD) effects in the ISM and the role of the magnetic fields.}

Magnetic fields play a central role in the formation, evolution, and stability of interstellar filaments. Numerical simulations of the ISM that include magnetic fields show a closer agreement with the observed filamentary organization of the ISM than purely hydrodynamic models \citep{federrath2010}. In these simulations, magnetic pressure and tension contribute to shaping the density structures and can provide support against gravitational collapse. Analytical models of magnetized filaments predict equilibrium configurations with shallower radial density profiles compared to gravitationally bound isothermal cylinders \citep{fiege2000}. At the same time numerical studies show that the interplay between turbulence, gravity, and magnetic fields is a decisive factor for the lifetime of the filaments. Thus, magnetic fields are not only involved in shaping filamentary structures but also in regulating their dynamical evolution.

Observational studies, in particular, \textit{Herschel} and \textit{Planck} observations, have revealed that filaments are ubiquitous in molecular clouds and are closely linked to star formation \citep{andre2010}. Moreover, filaments tend to exhibit substructure and fragmentation into velocity-coherent fibers \citep{hacar2013}, suggesting they are rather dynamically evolving objects. Magnetic fields are considered to be one of the main factors contributing to maintaining some coherence over parsec scales, while locally allowing fragmentation and collapse. Therefore, filaments can be viewed as intermittent (at the timescale of star formation) structures which are influenced by magnetic fields, turbulence, and gravity, and restructured by those in different proportions during the formation and evolution.

\textls[-15]{Relative orientation statistics provide a direct observational diagnostic of the coupling between gas and magnetic fields. This parameter allows us to probe the balance between turbulence, magnetic field and gravity \citep{soler2013,soler2017velac,alina2019}. Such studies provide a means of comparing theoretical predictions of MHD turbulence with observations and of assessing the role of the magnetic field in shaping the ISM, with implications on initial conditions of star formation. }

The identification and interpretation of filamentary structures remain an important component of the ISM studies as observational capabilities continue to expand.
All-sky dust emission polarization measurements from \textit{Planck} enabled studies of relative orientation statistics, while large area spectroscopic surveys of atomic and molecular gas have provided complementary kinematic information \citep{benbekhti2016,eden2020}. Ground-based (Pol-2 on JCMT \cite{ward-thompson2017}, HAWC+ on SOFIA \cite{harper2018}) and balloon-borne submillimetre polarimetric instruments (BLASTPol, ref. \cite{fissel2010,galitzki2014}, PILOT, ref. \cite{bernard2016}) resolve magnetic fields in individual clouds and filaments \cite{pattle2017,chuss2019,fissel2016,mangilli2019}. At smaller spatial scales, interferometric observations with ALMA probe magnetic fields in dense filaments and cores \citep{hull2017serpens,pillai2020}.

Forthcoming facilities and surveys promise to expand both the quantity and diversity of available datasets. New generation radio surveys with the Square Kilometre Array (SKA) and its precursors such as ASKAP (Australian SKA Pathfinder \cite{johnston2008}), MeerKAT \citep{jonas2009}, or LOFAR (low-frequency array \cite{vanhaarlem}), will provide and are already providing high-resolution spectroscopic maps and Faraday rotation measures of the magnetised ISM. These data sets will enable to further investigate connections between filamentary structures, kinematics, and magnetic fields \citep{johnston-hollitt2015,gaensler2015}. 
In addition, advances in millimeter and submillimetre instrumentation, including upgrades to existing facilities such as NIKA2 \citep{adam2018} and the development of new polarimetric instruments such as TolTEC \citep{toltec} and PRIMA (Polarized Radiation Imaging and monitoring Array ref. \cite{prima}) will improve the sensitivity and angular resolution. The large-scale future polarization missions like LiteBIRD \citep{hazumi2019} will provide high sensitivity all sky maps and will extend magnetic field studies to a wide range of environments, evolutionary stages and spatial scales. These facilities are expected to produce multi-scale, and multi-tracer datasets in which filamentary structures are a dominant morphological component. 

The already available survey datasets and other archival data has enabled the increasing use of automated and machine learning-based approaches for filament detection and characterisation \citep{koch2015,alina2022malefisenta,zavagno2023,umetaliev2025}. These methods are motivated by the need for scalability and reduced manual parameter tuning when analyzing hundreds to thousands of fields. While machine learning techniques offer clear advantages in handling big datasets and complexity, their performance remain closely linked to the assumptions embedded in classic approaches and preprocessing strategies \citep{micelotta2021,umetaliev2025}. 

In parallel with these observations-based studies, significant progress has been made in the theoretical description of MHD turbulence and filament formation. Numerical simulations explore wide parameter spaces in magnetic field strength, turbulence driving and self-gravity \citep{padoan2001,federrath2010,seifried2020}, and the existence of filaments was predicted in numerical simulations well before the availability of large-scale observations \citep{padoan2001,hennebelle2012,inoue2012}. However, the dependence of filament morphology, and of filament versus magnetic field alignment on the underlying MHD regime has been investigated in detail primarily in the last decade or so \citep{soler2013,planck2016-xxxv,chen2016,inoue2018}. These studies show that filament morphology and alignment are sensitive to the MHD regime. However, separating the effects of different regimes remains observationally difficult. Recent theoretical works emphasizes the role of magnetic field geometry and turbulence in cosmic ray transport and energy exchange \citep{zweibel2017,xu2020}. The filamentary associated with coherent magnetic fields could therefore also influence energy exchange in the ISM.

The study of filamentary structures significantly relies on the combined analysis of heterogeneous observational datasets. Integration of continuum and spectroscopic surveys, polarization measurements help to infer a broad range of physical properties, including density structure, kinematics, magnetic field orientation, and evolutionary stage \citep{malinen2016,cox2016,soler2019,alina2022monob}. Such studies involve datasets with different angular resolution, sensitivity, and spatial coverage. In this regard, the availability of archival data enables not only to design follow-up observations but also a systematic analysis of filamentary structures across a wide range of environments \citep{juvela2012}. As a result, filament identification becomes a critical step in the interpretation of the filament properties and their relation to magnetic fields and gas dynamics \citep{panopolou2017,schisano2020}.

In this review, we aim to provide a systematic overview of filament detection and characterization techniques. It is intended to help students and researchers who are beginning to work with data sets containing interstellar filaments. At the same time, the comparison of the different approaches may also be useful to more experienced researchers in assessing methodological biases and interpreting results obtained using different methods.

We group the existing approaches into methodological classes. For each of them, we discuss the scientific tasks for which they are commonly applied, as well as advantages and caveats. Particular attention is given to methods used in studies of relative orientation between the filaments and the magnetic field. With this, we attempt to clarify how methodological choices influence interpretation of filament properties. We also outline perspectives for the application of filament detection methods to large and diverse datasets that become available.

\section{Definition of a Filament}

\subsection{Morphological Definition}

Generally, a filament is understood as an elongated enhancement in column density or emission map relative to its local background. Such structures are typically characterized by a high aspect ratio, and a coherent ridge or crest \citep{andre2010,schisano2020}. 
Practically, the detection methods apply additional criteria, such as, among others, minimum spatial scales imposed by filtering procedures, persistence or significance thresholds, post-processing removal of short or low-significance branches \citep{sousbie2011,menshchikov2013,koch2015,clark2014,carriere2022fildreams}. 
For example, multi-scale filtering and segmentation techniques have been used to isolate elongated structures and estimate their widths in \textit{Herschel} continuum data \citep{menshchikov2013}. The resulting filament properties depend on the adopted spatial filtering, background subtraction, and crest identification procedures.  
Beyond contrast-based definitions, filamentary structures have also been identified through their coherent linear anisotropy. For example, elongated HI features identified in diffuse atomic gas \citep{clark2014} and similar orientation-based studies \citep{juvela2016,carriere2022fildreams} treat filaments as regions exhibiting preferred direction at a given spatial scale. In such cases, a filament is characterised as a directional pattern within the intensity field.

These examples illustrate that the observational concept of a filament is not uniquely defined. Depending on the tracer and the adopted analysis approach, filaments may be identified through contrast, connectivity, or directional coherence. The corresponding choices affect which elongated structures are retained as filaments in a given dataset.

\begin{figure}[H]
\includegraphics[width=0.3\linewidth,trim=30 10 80 10]{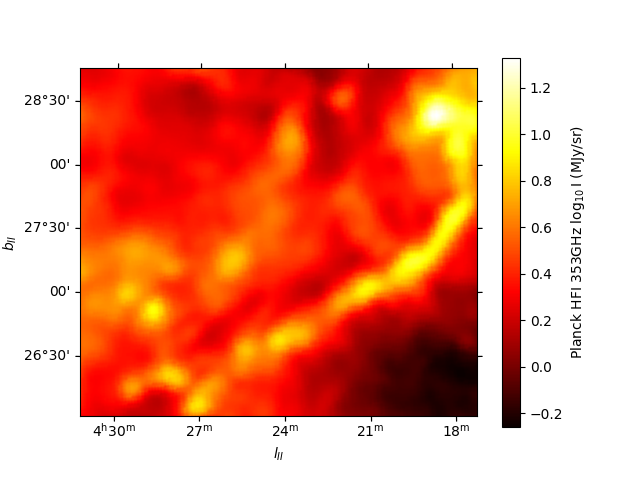}
\includegraphics[width = 0.3\linewidth, trim=30 10 80 10]{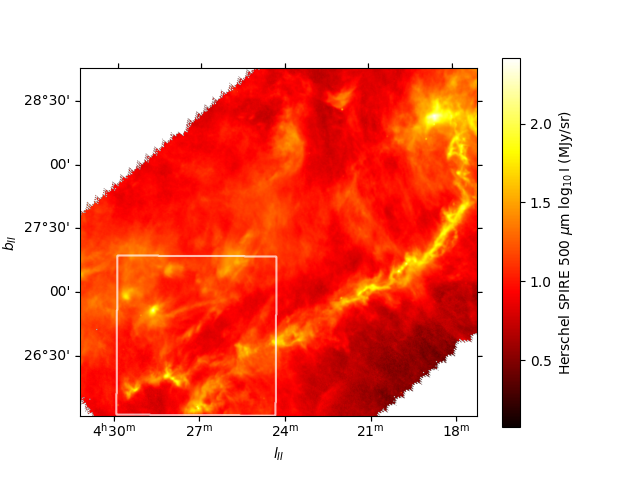}

    \caption{A part of Taurus molecular cloud. Left: \textit{Planck} 353 GHz with 7' resolution, right: \textit{Herschel} 500 $\upmu$m with 36' resolution. The white square delimits the region used further in Figure~\ref{fig:comparison}.}
    \label{fig:taurus}
\end{figure}

In addition, the apparent morphology of a filament is scale-dependent. As an example, the Taurus molecular cloud appears notably different when observed at different angular resolutions and different instruments. Large-scale \textit{Planck} map emphasizes a dominant elongated structure in the left panel Figure~\ref{fig:taurus}, while higher-resolution \textit{Herschel} observations resolve a complex network of sub-filaments and striations. Thus, the identification of a structure as a single filament or as a filamentary complex depends on the spatial scale and resolution at which it is observed.

\vspace{-12pt}

\vspace{-6pt}

\begin{figure}[H]
    \includegraphics[width=0.6\linewidth]{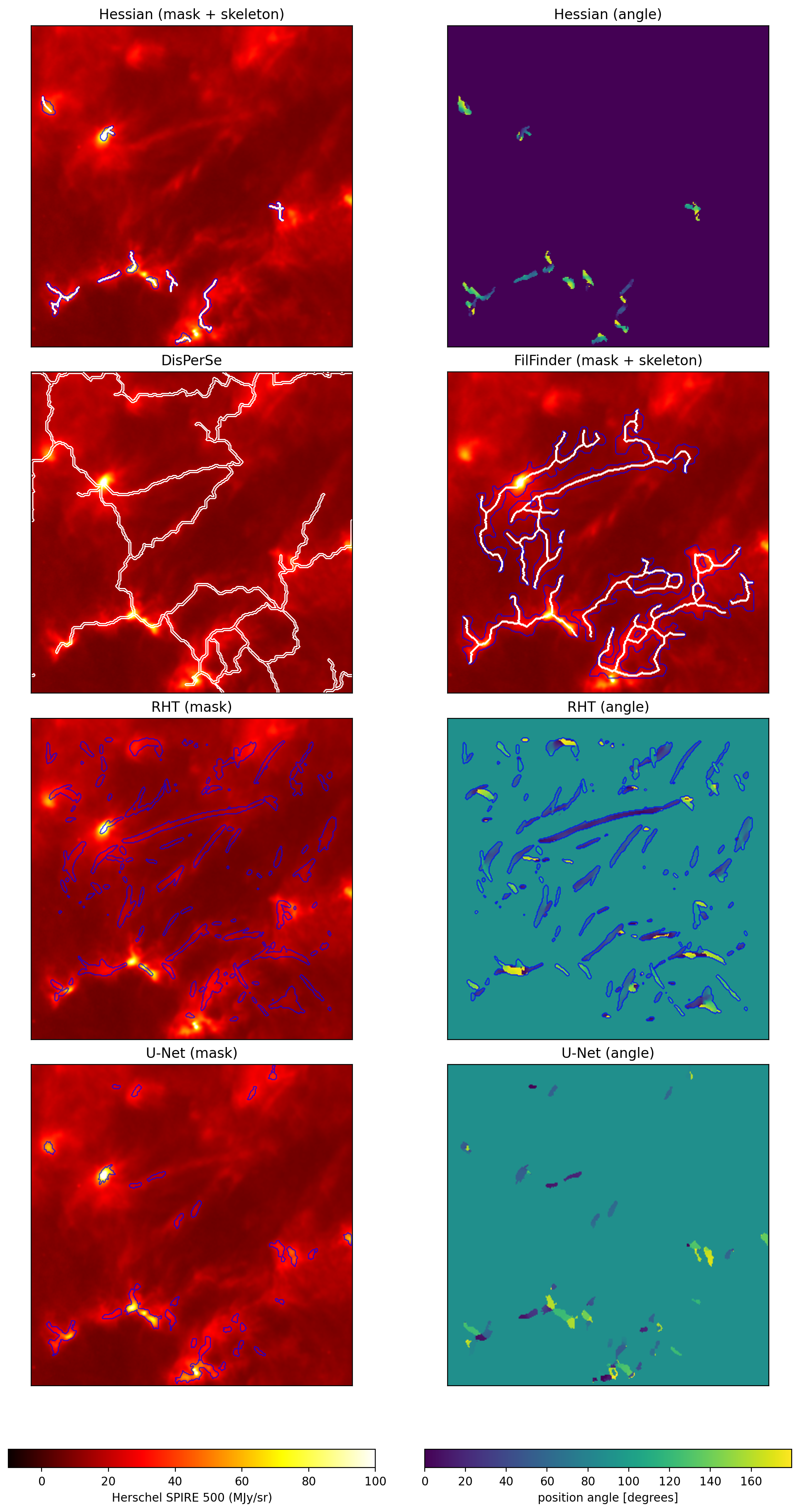}
    \caption{Results 
 of the different filament detection methods applied to the same map, shown in the white rectangle in Figure~\ref{fig:taurus}. First row: intensity map with the structures identified using Hessian matrix-based approach and the derived position angle. Second row: intensity maps overlaid with the features detected by {\disperse} (left) and {\filfinder} (right). Note that the skeleton of {\filfinder} was obtained at the post-processing step. Third row: intensity map overlaid with the contours of the RHT result (left) and the derived orientation angles (right). Fourth row: intensity map overlaid with the contours of the filaments obtained at the pre-processing (left) and the position angles derived using the ML-based approach \citep{umetaliev2025}.}
    \label{fig:comparison}
\end{figure}

\subsection{Projection Effects}
\label{sec:projection}

A fundamental complication arises from the fact that most filament detections are performed in two-dimensional projected maps. The observed structure may represent:
\begin{itemize}
    \item a coherent three-dimensional density enhancement;
    \item the superposition of multiple unrelated structures along the line of sight;
    \item a velocity-coherent feature identifiable only in the position-position-velocity space;
    \item the projection of an inclined sheet-like structure.
\end{itemize}

In the latter case, a geometrically thin sheet viewed edge-on can appear as a filament in projection \citep{heitsch2013,tritsis2018tassis}.

Spectroscopic studies have shown that filamentary structures in dust emission can decompose into multiple velocity-coherent substructures when analyzed in molecular line data \citep{hacar2013,hacar2018}. 

Projection effects become particularly important in the studies of hub-filament systems, where either many elongated structures appear to converge toward a central dense clump or toward a main filament. In projected maps, such configurations may suggest a coherent network feeding material into a hub or main filament. 
However, without full three-dimensional information, it remains difficult to determine whether these filaments are physically connected or simply superposed along the line of sight. Spectroscopic observations often reveal complex velocity structures that can either support or challenge the apparent morphology \citep{hacar2013,hacar2018}.
In observations, the true direction of gas flow along filaments is not directly observable. Although gradients in velocity maps may be interpreted as accretion toward hubs, filaments or cores, projection and inclination effects can significantly alter those interpretations. Therefore hub-filament systems are a clear example of how projection effects and projected velocity ambiguities can complicate filament identification and interpretation.  

Projection effects can also influence measurements of relative orientations with respect to magnetic fields, since both filament extent and magnetic fields are derived from projected observables. The plane-of-the-sky magnetic field derived from polarization measurements,  in starlight extinction and in thermal emission, represents only a component of the full three-dimensional field geometry.

Several studies have attempted to overcome this limitation by reconstructing the three-dimensional structure of filamentary systems and magnetic fields. Statistical techniques based on anisotropies of magnetized turbulence, such as the velocity gradient technique \citep{lazarian2018vgt,hu2023lazarian,hu2024lazarian}, provide indirect constraints on the three-dimensional magnetic field orientation using spectroscopic observations. In parallel, analyses of filamentary structures in the diffuse atomic medium have shown that HI filaments are often aligned with the plane-of-the-sky magnetic field \citep{clark2019,kalberla2016}, suggesting that filamentary structures themselves can act as tracers of magnetic field geometry. While these approaches do not yield a direct reconstruction of the full three-dimensional configuration, they offer valuable statistical and observational constraints on the underlying structure of the magnetized ISM. More importantly, new techniques employing Faraday rotation measures along with Galactic magnetic field models have been developed to reconstruct the three-dimensional structure of molecular clouds (\cite{tahani2022} and references therein).

Numerical modeling offers another means for understanding projection effects. MHD simulations combined with radiative transfer modeling \citep{seifried2015,juvela2019} allow direct comparison between intrinsic three-dimensional structures and their projected observational counterparts. These studies demonstrate that both filament morphology and inferred magnetic field orientation depend sensitively on viewing angle and line-of-sight integration.

\subsection{Physical Filaments Versus Observational Filaments}

Projection effects discussed in the previous section represent one source of ambiguity. However, even when the projection effects are not dominant, a morphological filament is not necessarily a gravitationally bound, dynamically coherent structure or a magnetically supported structure.

Filaments are often discussed as physical structures shaped by turbulence, gravity, and magnetic fields, while observationally, they are identified as elongated intensity or column density enhancements in projected maps.
In molecular clouds, some filaments show evidence of velocity coherence and internal substructure in position-position-velocity (PPV) space \citep{hacar2013,hacar2018}. In other cases, elongated features identified in continuum emission may arise from superposition effects.

These distinctions highlight that the term ``filament'' can refer either to an observationally defined elongated structure or to a physical structure with specific dynamic properties. The relationship between these two interpretations is not always straightforward. A filament identified through morphological criteria may or may not correspond to a coherent physical entity in three dimensions.

We stress that recognizing this distinction is important for interpreting filament properties or alignment with magnetic fields. Scientific interpretation depends on how the filament has been defined and extracted from the data.

\subsection{Observational and Methodological Filaments}
The absence of a universally adopted definition means that filaments should be seen as method-dependent constructs derived from observational data. A filament identified through topological skeleton extraction is not necessarily identical to a filament identified via linear anisotropy-based method. The resulting structures may differ in length, width, spatial extent.

This difference does not criticize the validity of any of the methods. It rather illustrates the importance of aligning the detection technique with the underlying physical question. As an example, the identification of a characteristic filament width of approximately 0.1 pc from \textit{Herschel} observations has played a central role in shaping the current understanding of filament structure. While subsequent studies have reported similar results in many regions, other showed that the inferred width distribution can depend on the specific filament extraction method \citep{panopolou2017}, angular resolution or pre- or post-processing. This example shows that interpretations are intrinsically linked to instrumental characteristics (such as the sensitivity and angular resolution of the telescopes) and to methodological choices.

\section{Classification of Filament Detection Methods}
\subsection{Mathematical Formulation}
Although filament identification in three-dimensional data is the subject of ongoing research, we restrict the formalism exposed below to the two-dimensional case.

Let $I(x,y)$ denote a scalar field defined over a two-dimensional domain. $I(x,y)$ can be an intensity map, column density, or another observable. Filament detection methods can be understood as operators acting on the scalar field:
\begin{equation}
    \mathcal{F}: I(x,y) \rightarrow \mathcal{O}(x,y) \, ,
\end{equation}
where $\mathcal{O}$ may include quantities such as skeleton, local orientation, characteristic scale, or significance measures. Although implemented differently, most filament identification algorithms can be grouped according to the mathematical meaning of the operator $\mathcal{F}$.

Below we propose a classification of the detection methods, and it is not intended as an exhaustive survey of all existing techniques. We also note that almost all filament detection methods implicitly or explicitly introduce a characteristic spatial scale through smoothing, filtering or windowing operators. 

\subsection{Local Differential Operators}
One class of filament detection methods is built upon differential properties of the scalar field. 
The corresponding  fundamental operators are a simple gradient ($\frac{\partial I}{\partial x}$, $\frac{\partial I}{\partial y}$) or the Hessian matrix:
\begin{equation}
    H(I)(x,y)=
\begin{pmatrix}
\frac{\partial^2 I}{\partial x^2} & \frac{\partial^2 I}{\partial x\,\partial y} \\
\frac{\partial^2 I}{\partial y\,\partial x} & \frac{\partial^2 I}{\partial y^2}
\end{pmatrix}.
\end{equation}

Examples include Hessian-based extraction methods \citep{molinari2010,schisano2014} and gradient-based orientation studies (HRO \cite{soler2013}).

In Hessian-based approaches, filaments are associated with anisotropic curvature: one eigenvalue of $\mathcal{H}(I_{l})$ is significantly negative, corresponding to the direction perpendicular to the filament, while the second eigenvalue is relatively small. Here $I_l$ denotes the map filtered at scale $l$. The local filament orientation is given by the eigenvector corresponding to the smallest curvature direction. This class of methods is intrinsically local, linear with respect to derivatives and explicitly scale dependent through smoothing or filtering.

\subsection{Template and Model-Matching Methods}
\textls[-15]{A second major class of filament identification methods is based on matching the observed data to a predefined geometric template. In these approaches, the orientation of a filament is inferred by evaluating how well the data agree with the template at different rotation angles.
In contrast to differential methods, which rely on local curvature, template-based approaches quantify structural coherence by direct comparison with the reference shape.}

Among the existing methods, we differentiate between linear convlution-based and geometric multi-scale approaches.

\subsubsection{Linear Convolution-Based Template Matching}
Template Matching (TM \cite{juvela2016}) is an example of classical template matching applied to filament detection. 
Let denote $\mathcal{T}_{\theta}(x,y)$ as a geometrical template rotated by an angle $\theta$. The corresponding operation on an intensity map $I$ is a convolution with the rotated template: 
\begin{equation}
    S(x,y,\theta) = (I \ast T_{\theta})(x,y).
\end{equation}
The 
 local orientation is then determined as the angle which maximizes the result of the convolution. 

\subsubsection{Hough-Type Transforms}
The Rolling Hough Transform (RHT \cite{clark2014}) follows a similar approach but performs local transformation from spatial $(x,y)$ coordinates to orientation space. In the classical Hough transformation, linear features are represented in $(\rho,\theta)$ space, where $\theta$ is the angle and $\rho$ is the distance from the origin. RHT uses a sliding window centered at each pixel (so $\rho$ is equal to $0$) and  evaluates the degree to which pixels are aligned along a given direction $\theta$. 
For every pixel, the final orientation is evaluated from the statistical analysis of the angular distribution. 

\textls[-15]{Although RHT is technically different from simple convolution, it is based on the fundamental idea of evaluating orientation relative to an elongated template (typically a bar).}

\subsubsection{Geometric Multi-Scale Matching}
FilDReaMS \cite{carriere2022fildreams} is another template-based method, with additional geometric and statistical elements, that allows for a multi-scale filament detection. 
Unlike classical template matching and the RHT, where the characteristic scale is fixed by the template size or window diameter, FilDReaMS iterates over a range of bar widths and determines the most significant one through Monte Carlo comparisons. 

\subsubsection{Comparison of Template-Based Methods}

Although classical template matching, the RHT, and FilDReaMS are based on the common principle of directional coherence with a reference structure, they differ in mathematical formulation and scale treatment.

TM operates directly on the intensity field and computes a convolution-based response. The characteristic scale is fixed by the template dimension. As a result, the method is inherently single-scale unless repeated with multiple templates. One of the advantages of the method is in its flexibility in using different shapes: it allows using alternative geometries such as curved or profile-based shapes. Although similar extensions could be implemented in other methods, TM or linear template matching methods in general, allows straightforward implementation of this modification.

The RHT also evaluates directional coherence but operates locally in the orientation space. The effective scale is determined by the window diameter, which in practice translates to a rotating bar length and width. The output is the accumulated evidence of colinearity rather than the result of direct convolution. Similar to TM, the scale is thus predefined and multi-scale analysis requires repeating the procedure for different scales.

FilDReaMS extends the template-based philosophy by explicitly scanning over a range of bar widths and selecting statistically most significant configuration at different angles. In addition, it provides an estimate of characteristic width and incorporates Monte Carlo-based validation. This approach enables to address the size bias issue inherent to single-scale methods.

These differences directly impact computational cost. In single-scale template matching, runtime scales with the number of tested orientations, and in multi-scale approach - additionally with the number of template sizes. Scale detection also influences parameter dependence. In single-scale approaches, detection sensitivity and orientation stability depend on the chosen template size or window diameter. In multi-scale approaches, results depend on the width range and statistical thresholds. Consequently, appropriate parameter selection requires considering the scientific objectives and the specific characteristic of the intensity map of interest.  

\subsection{Topological and Skeleton-Based Methods}
A third class of filament identification methods is based on topological analysis and skeleton extraction. These methods focus on connectivity and persistence of structures in the map. Unlike the differential and template-based, the detection is formulated at the level of extended objects rather than individual pixels.

DisPerSE \citep{sousbie2011} is based on discrete Morse theory, which analyzes the topology of a scalar field through its critical points and the gradient flow connecting them. In practice, DisPerSE identifies maxima and saddle points in the map and traces integral lines between them to reconstruct the underlying structures. A persistence criterion is then applied to remove features that may be arising from noise. The resulting detection is attributed to the topologically significant connected structures.

FilFinder \citep{koch2015} follows a similar morphology-based approach. After mask construction and thresholding, medial-axis skeletonization is applied to extract filament spines. Similarly, multi-scale structure separation tools such as \textit{getsf} \citep{menshchikov2013,menshchikov2021} isolate filamentary components prior to skeleton extraction.

\textls[-15]{Topological and skeleton-based methods are particularly well-suited for identifying filament crests and reconstructing their connectivity. The primary output is one-dimensional spine or network representation of the filamentary structure. However, these approaches do not directly provide pixel-wise orientation angles or a continuous mask describing the spatial extent of a filament. Additional operations, such as a tangent estimation along the skeleton or dilation of the spine to recover the filament width, should be performed if necessary.}

\subsection{Machine Learning-Based Methods}
Machine learning-based (ML-based) methods differ fundamentally from deterministic methods described above. Here, the operator mapping the input map to filament properties is not analytically defined but learned from data. These approaches train a parametric function to approximate the desired mapping from the scalar field $I(x,y)$ to a target representation. The operator can be written as
\begin{equation}
    \mathcal{F}[I] = f_{\mathrm{w}}(I)\, ,
\end{equation}
where $f_{\mathrm{w}}$ is a nonlinear function parametrised by weights ${\mathrm{w}}$, learned through optimization with respect to a loss function defined on labeled data.

In ML-based methods \citep{alina2022malefisenta,zavagno2023,umetaliev2025}, detection and identification are not necessarily separate stages. Depending on the objective, a neural network (NN), which is most commonly used ML technique in filament studies, may simultaneously produce a filament mask or an orientation field. The behavior of the operator is thus determined by the training dataset and chosen architecture rather than by explicit analytical criteria. However, the analytical criteria lie intrinsically within the labeled dataset.

In one of the methods \citep{zavagno2023} , U-Net architecture \citep{ronneberger_u-net_2015} was used to perform binary filament segmentation, training the network on masks derived from a Hessian-based detector. In this case, the model learns to reproduce curvature-based detection and provides filament detection. 

The U-Net architecture has been also used for a multi-class semantic segmentation to predict a pixel-wise orientation field, based on the training on pixel-wise orientation synthetic two-dimensional maps \citep{umetaliev2025}. This approach directly targeted orientation identification rather than approximating an existing deterministic method. 

Deterministic methods described in previous sections, offer transparency as each step is a well-defined mathematical operation. However, they often require explicit parameter tuning. ML-based methods reduce parameter tuning for each individual map but are sensitive to the statistical properties of the training data and preprocessing pipeline. Thus, the choice between deterministic and ML-based approaches depend on the scientific objective: analytical control or data-driven adaptability.

\section{Methods Summary and Application to Real Data}

Table~\ref{tab:1} provides a summary of the methods discussed in this review. 
To visually demonstrate the differences between the different methods, we apply several representative identification methods to the same observational field shown in the white rectangle in Figure~\ref{fig:taurus}. The resulting comparison presented in Figure~\ref{fig:comparison} has only a qualitative character. It is worth noting that for every method, the output would vary depending on parameters chosen. However, we believe such a comparison can provide the reader with a global picture. We also stress that the goal is not to establish a performance ranking but to highlight how methodological choices influence the resulting filament representation.

Parameters that we used to produce the result shown in Figure~\ref{fig:comparison} are reported in Table~\ref{tab:params}. For visualization purposes, additional post-processing was applied to some methods to facilitate comparison of the identified structures. The Hessian-based method produces eigenvalues (filament "strength" and curvature). So, to visualize the detected structure, we derived a binary mask at the 90th percentile threshold and removed small regions and holes. For FilFinder, we have also derived the skeleton from the mask. 

\begin{table}[H]
\tablesize{\small}
\caption{Representative 
filament identification methods and their native outputs.}

\newcolumntype{C}{>{\centering\arraybackslash}X}
\begin{tabularx}{\textwidth}{LLL}
\toprule
\textbf{Method} & \textbf{Class} & \textbf{Native Output} \\ 
\midrule
Hessian-based & Differential & Orientation field/mask \\
Template matching \citep{juvela2016} & Template-based & Orientation field \\
RHT \citep{clark2014} & Template-based & Orientation distribution \\
FilDReaMS \citep{carriere2022fildreams} & Template-based & Orientation + width \\
DisPerSE \citep{sousbie2011} & Topological & Skeleton \\
FilFinder \citep{koch2015} & Morphological & Mask \\
\textit{getsf} \citep{menshchikov2013} & Multi-scale decomposition & Mask + skeleton \\
U-Net mask \citep{zavagno2023} & Learning-based & Filament mask \\
U-Net angle \citep{umetaliev2025} & Learning-based & Orientation field \\
\bottomrule
\end{tabularx}
\label{tab:1}
\end{table}

\textls[-25]{Figure~\ref{fig:comparison} shows that the different methods emphasize different aspects of the filamentary structures seen in the map. The Hessian-based approach highlights only the most prominent ridges, resulting in a small number of structures associated with the brightest crests. DisPerSE trances larger network of filaments, including many faint and extended structures. FilFinder produces an intermediate representation, capturing broader structures, while the post-processed skeleton follows the brightest ridges. 
RHT detects both bright and faint elongated structures resulting in many short or long quasi-straight structures. In addition, it provides the position angles associated with each pixel of the identified filaments. The U-Net-based method tends to emphasize the most prominent filaments in the field, does not necessarily requires long aspect ratios and gives an estimate of position angles along these structures.}

\begin{table}[H]
\tablesize{\small}
\caption{Parameters used for the illustrative comparison shown in Figure~\ref{fig:comparison}. The input map size is $321 \times 321$ pixels.}

\newcolumntype{C}{>{\centering\arraybackslash}X}
\begin{tabularx}{\textwidth}{lL}
\toprule

\textbf{Method} & \textbf{Parameters Used} \\ 
\midrule
Hessian-based & $\sigma = 1.5$, $\beta = 0.5$, threshold percentile = 90, hole size = 80 px, min\_size = 80 px \\
\midrule
RHT \citep{clark2014} & Kernel size = 27 px, bar width = 6 px, smoothing = 6 px \\
\midrule
FilFinder \citep{koch2015} & Beamwidth $\approx 36''$; minimum branch length = 0.1 pc (at a distance of 140 pc); minimum skeleton length = 0.05 pc \\
\midrule
DisPerSE \citep{sousbie2011} & Default persistence and noise parameters \\
\midrule
U-Net \citep{umetaliev2025} & Herschel preprocessing pipeline as described in the original paper \\
\bottomrule
\end{tabularx}
\label{tab:params}
\end{table}

\section{Discussion}

\subsection{Parameter Dependence and Projection in Connection to Hub-Filament Studies}
\label{sec:hfs}

Filament detection methods are intrinsically parameter-dependent. In skeleton-based and topological approaches such as DisPerSE, FilFinder and \textit{getsf}, persistence and significance criteria determine which features are considered as meaningful structures. Even moderate variations in the input parameters can modify the outputs: filaments continuity, branching, and length distributions. 
This sensitivity reflects the hierarchical nature of interstellar filamentary structures rather than methodological issues. Different parameter settings provide insights into different structure levels within the same cloud. 

The skeleton-based and topology-based methods are advantageous for characterizing the filamentary networks' topology. These methods allow identification of junction points. Hub-filament systems have been analyzed using such skeleton reconstructions to quantify the convergence of multiple filaments toward dense star-forming clumps.

As already discussed in Section~\ref{sec:projection}, one of the fundamental challenges in studying hub-filament systems arises from the projection effects of the three-dimensional structures onto two-dimensional observational maps and the confusion along the line of sight of physically distant structures. This superposition can lead to apparent junctions. 
Spectroscopic observations are used to overcome this ambiguity by providing line of sight velocity information, e.g., PPV cubes were used to identify velocity-coherent filaments and provide hints on the gas flows toward a hub in G33.92 + 0.11 OB cluster-forming region \citep{liu2012}, or later in the study of Mon R2 region \citep{trevino-morales2019}. In fact, the interplay between gravity, velocity gradients, and magnetic fields can only be interpreted consistently when kinematic information is taken into account \citep{wang2020}. 

Projection effects influence the inferred geometry of filaments, as the apparent aspect ratio and curvature depend on viewing angle. Since collapse timescales are strongly geometry-dependent \citep{pon2011,pon2012}, misinterpretation of intrinsic structure may lead to inaccurate conclusions about dynamical evolution.

Magnetic field studies are also affected by projection effects, e.g., ref. \cite{arzoumanian2021} in the study of NGC6334 showed that dust polarization reveals organized magnetic structures associated with hub-filament systems, yet the observed plane-of-the sky magnetic field orientation represents only a projected component of full 3D geometry. 

These studies highlight that reliable characterization of hub-filament systems requires combination of continuum morphology with spectroscopic and, ideally, polarization data. Although projection effects cannot be eliminated directly, multi-component diagnostics can significantly improve physical interpretations.

Numerical modeling provides a powerful tool for understanding projection effects in hub-filament systems. Numerical simulations of molecular clouds and filament formation (e.g., refs. \cite{nagai1998,inutsuka2015}) provide insight into the intrinsic three-dimensional structure of the clouds. Polarization modeling is another efficient probe of projection effects. Radiative transfer calculation applied to MHD simulations of turbulent clouds \citep{seifried2015,juvela2019} show that both polarization fraction and angle depend sensitively on viewing angle and line-of-sight tangling. 
Such modeling, especially when combined with synthetic observations, can provide us with a unique tool of quantifying how projection modifies the inferred morphology and field orientation of filamentary networks.

\subsection{Implications on the Relative Orientation Analysis}
Relative orientation between filaments and magnetic fields has become a key observational probe of dynamical state of filamentary clouds. 
Observational studies use statistical tools, such as the Histogram of Relative Orientations (HRO, ref. \citep{soler2013}) or analog parameters, to quantify whether the filaments tend to align parallel or perpendicular to the local magnetic field \citep{planck2016-xxxv,jow2018,alina2019,fissel2019,soler2019}. These alignments are then interpreted in terms of relative importance of turbulence, gravity, and magnetic fields \citep{chen2016,soler2017velac,seifried2020}.

MHD simulations show that relative orientation depends on the degree of magnetization and on the interplay between turbulence, gravity and compressive motions. Sub-Alfv\'enic turbulence has the tendency of producing density structures preferentially aligned with magnetic field, while gravitational contraction can lead to structures that are mostly perpendicular to the field at higher densities. In addition, shock compression, anisotropic turbulence driving and local feedback can also generate density enhancements. In this case, the relative orientation with respect to the magnetic field depends on the pre-existing field configuration and on the geometry of the shock \citep{abe2021}. Thus, the resulting alignment is sensitive to the Mach number, the magnetization level, the evolutionary stage of the region and a local dynamical regime. Such a diversity of physical configurations indicates that the alignment trends, although very informative, must also be interpreted with care.

Methodological factors further influence alignment statistics. Determination of filament orientation depends on the extraction algorithm, smoothing and masking. The latter defines which structural scale is being analyzed. Changes in spatial resolution or density threshold can modify the resulting angle distribution and can shift the apparent transition density. These dependencies highlight that alignment measurements are intrinsically scale-sensitive and must be interpreted in the context of the defined structural components, e.g., filaments, clumps, sub-pc filaments, etc.

\subsection{Radial Profiles of Magnetized Filaments}
The radial (transverse to the longest dimention) density profile of filaments provide one of the most direct constraints on their physical origin and serve as a test bench for different theories.
Observationally, filament profiles are characterized by a Plummer-like density function:
\begin{equation}
    \rho(r) = \rho_c {\left[1 + \left(\frac{r}{R_{\mathrm{flat}}}\right)^2 \right]^{-p/2}} \, .
\end{equation}
Here, $\rho_c$ is the central density, $R_{\mathrm{flat}}$ is the radius of the inner flat region, and $p$ is a power-law index. 
An isothermal self-gravitating cylinder yields a profile of form $r^{-4}$ \citep{ostriker1964}, whereas magnetized configurations can produce shallower slopes \citep{fiege2000}. \textit{Herschel} observations of filaments in IC5146 showed their profiles described by $p\simeq 2$ and inner width of 0.1 pc \cite{arzoumanian2011}. Subsequent \textit{Herschel} surveys also used radial fitting to explore filament stability and mass per unit length distributions \citep{schisano2014}. 

\textls[-15]{More recently, large-scale molecular line surveys such as SEDIGISM \citep{schuller2017} have extended radial profiles to Galactic plane environments with velocity information. The SEDIGISM survey revealed a wide diversity of filament widths, line masses and dynamical states \citep{duarte-cabral2021}. These results suggest that filament radial structures vary systematically with environment, evolutionary stage, and do not conform to a single characteristic width or equilibrium solution.}

\textls[-15]{Simulations of turbulent magnetized flows show that filament profiles can arise naturally from anisotropic compression or converging flows without demanding hydrostatic equilibrium.
Numerical studies showed that filament profiles similar to observed Plummer-like shapes can arise in supersonic tubrulence, gravitational contraction along magnetic field lines, and through continuous accretion toward the densest filament from the striations \cite{smith2014,clarke2016,federrath2016}. }

Thus, Plummer-like radial profiles are not a unique diagnostics of the dynamical state of a filament. 
Here the connection to relative orientation with respect to magnetic fields may provide an additional insight. If radial profile and alignment both arise from anisotropic MHD processes, their combined analysis can better constrain the underlying dynamical regime.
In this case, both observables should be interpreted within the broader context of dynamically evolving magnetized flows in molecular clouds.

In terms of the filament detection methods, the inference of radial profiles depends on how the filament crest and boundaries are defined. In the studies cited above, profiles were derived by tracing a spine with skeleton-based or topological method, computing perpendicular cuts, averaging along the crest and performing background subtraction before fitting. While radial fitting provides a powerful parametric description, comparisons between regions and surveys require careful consideration of the detection method and the effects of telescope sensitivity and resolution.

\section{Perspectives}

Filament properties are used to infer physical processes in the interstellar medium, taking as starting point the filament widths, radial profiles, or relative orientation with respect to magnetic fields. Consequently, filament detection methods have seen rapid development in the last few decades.

We show here that filament identification is not a uniquely defined operation. Different methods apply different mathematical operations, and as a result, the structures identified by different techniques can differ significantly even when applied to the same map. However, this should not be interpreted as a limitation of methods. Interstellar clouds are intrinsically hierarchical, spanning from several parsecs to tens of thousands of au. In addition, telescope resolution and sensitivity provides access to different parts of filamentary networks. On top of that, the filamentary network is partly determined by the applied method. As a consequence, statistical quantities derived from these structures also depend on the chosen detection method. 

A well-known example of using filament detection methods for inferring physical properties concerns the study of the relative orientation between filaments and magnetic fields. 
It remains one of the most powerful statistical tools for probing the interplay between magnetic fields and density structures in the ISM, in particular when combined with complementary analyses of the velocity information, mass derivation, density profiles and numerical modeling.
Observational studies have revealed systematic trends, often interpreted as signatures of magnetized turbulence and the interplay between turbulence, gravity, and magnetic field. In particular, large-scale analyses of dust polarization maps using the \textit{Planck} satellite data have shown that the relative orientation between filaments and the magnetic field can change with density or environment \citep{planck2015-XXXII,soler2017velac,alina2019}. Such results provide observational constraints on theoretical models of MHD turbulence and the degree of magnetization of molecular clouds. In this context, reliable determination of filament orientation becomes a key step for statistical analyses of the anisotropic structure of the interstellar medium.

The comparison presented in this work highlights that measurement of orientation statistics can depend on the extraction method and on the spatial scales emphasized by the algorithm. Methods that identify only the brightest ridges may trace structures associated with dense filaments and gravitational contraction, while methods that are sensitive to more diffuse emission may capture larger scale anisotropic patterns shaped by turbulent flows and magnetic fields. As a result, orientation statistics derived from different filament detection methods may probe different physical regimes of the ISM. Understanding these methodological effects is therefore important when interpreting the results of studies involving magnetized turbulence.
One of the promising direction can be the combination of different approaches. 
For example, topological methods can be used to identify filaments while machine learning-based methods can be employed to derive orientation angles. Thus, the transparency and interpretability of the conventional detection methods can be combined with the speed and scalability of machine learning-based techniques.

Beyond the detection methods discussed here, there exist additional approaches developed to analyze interstellar clouds. SCIMES \citep{colombo2015} algorithm applies spectral clustering techniques to spectroscopic observation data in order to identify coherent velocity and spatial structures within molecular cloud complexes. Wavelet-based analyses can provide another means of studying filamentary structures across multiple spatial scales. These approaches are useful for separating structures associated with different physical scales and may complement filament detection methods.

Another important direction concerns the reconstruction of the three-dimensional structure of the interstellar medium. Most filament detection techniques and magnetic field relative orientation analyses operate on projected two-dimensional maps. However, several studies have explored the possibility of considering filaments themselves as a proxy for magnetic field geometry. HI filaments have been shown as aligned with plane-of-sky  magnetic field orientation across the sky \citep{clark2019}, although this tendency is valid over very large regions in the sky and should be used with caution \citep{alina2023}. Similar trends have also been reported in the cold neutral medium \citep{kalberla2016}. The theory of the anisotropic MHD turbulence predicts that density and velocity structures tend to have preferential orientations with respect to the magnetic field \citep{goldreich1995}. These studies suggest that filaments can be seen as tracers of the magnetic field geometry and may threfore help to constrain the three-dimensional structure of the magnetized ISM.

Recent developments in filament detection algorithms aim to extend the methods to three-dimensional datasets. For example, the FilFinder 3D algorithm is an extension of FilFinder to simulation data cubes and position-position-velocity spectroscopic data (used, e.g., in \cite{zucker2021}). 
In parallel, several recent works have explored the possibility of inferring three-dimensional magnetic field properties using statistical and machine learning approaches based on anisotropies of the turbulent ISM, applied to molecular spectroscopic data and synchrotron observations \citep{hu2020,hu2023lazarian,hu2024,hu2025a,hu2025b}. 

Future studies of filamentary structures in the ISM will likely rely on a combination of complementary approaches. Although two-dimensional filament detection will remain essential, the three-dimensional reconstruction techniques may help connect projected structures with the underlying physical properties of the gas and magnetic field. Combination of these approaches may provide a more complete picture of how turbulence, magnetic fields, and gravity shape the anistropic structure of the interstellar medium.


\vspace{6pt}

\funding{This work was supported by the Faculty Development Competitive Research Grant Program of Nazarbayev University No. 201223FD8821. During the preparation of this manuscript, the authors used ChatGPT version 5.3 for the purpose of formatting the bibliographic references list. The author has reviewed and edited the output and takes full responsibility for the content of this publication. The author thanks Baurzhan Kumarioldanov and Temirkul Umetaliev for running {\disperse} and {\filfinder} algorithms, and the U-Net model on the data employed in this review.}

\dataavailability{Data used in this review have been retrieved from ESASky, developed by the ESAC Science Data Centre (ESDC) team and maintained alongside other ESA science mission's archives at ESA's European Space Astronomy Centre (ESAC, Madrid, Spain). \\ The authors declare no conflict of interest}



\abbreviations{Abbreviations}{
The following abbreviations are used in this manuscript:\\
\noindent
\begin{tabular}{@{}ll}
ISM & interstellar medium\\
{\disperse} & Discrete Persistent Structures
Extractor \\
RHT & Rolling Hough Transform\\
TM & Template Matching\\
{\fildreams} & Filament Detection \& Reconstruction at Multiple Scales\\
MHD & magnetohydrodynamic \\
JCMT & James Clerk Maxwell Telescope \\
HAWC+ & High-resolution Airborne Wide-band Camera Plus\\
SOFIA & Stratospheric Observatory for Infrared Astronomy\\
BLASTPol & Balloon-borne Large Aperture Submillimeter Telescope for Polarimetry\\
PILOT & Polarized Instrument for Long wavelength Observation of the Tenuous interstellar  \\
&  medium\\
ALMA & Atacama Large Millimeter Array\\
SKA & Square Kilometre Array\\
ASKAP & Australian SKA Pathfinder \\
NIKA & Néel IRAM KID Array\\
PRIMA & Polarized Radiation Imaging and monitoring Array \\
LiteBIRD & Lite (Light) satellite for the study of B-mode polarization and Inflation from cosmic \\
& background Radiation Detection\\
PPV & position-position-velocity\\
\end{tabular}
}


\begin{adjustwidth}{-\extralength}{0cm}

\reftitle{References}

\PublishersNote{}
\end{adjustwidth}
\end{document}